# Magnetoplasmonic Nanopore Lensing for Enhanced Optical Readout and Controlled Translocation

*Nageswar Reddy sanamreddy*[1,2], *Paolo Vavassori*[1,3]

[1]CIC nanoGUNE BRTA, Tolosa Hiribidea, 76, E-20018 Donostia-San Sebastian, Spain

[2]Department of Physics, University of the Basque Country (UPV/EHU), E-20018 Donostia-San Sebastian, Spain

[3]IKERBASQUE, Basque Foundation for Science, Plaza Euskadi, 5, E-48009 Bilbao, Spain

Email: nr.sanamreddy@nanogune.eu; p.vavassori@nanogune.eu

**ABSTRACT**

Plasmonic nanopores hold a significant promise for molecular sequencing, but their sensitivity and temporal resolution are constrained by limited signal strength and rapid translocation of molecules through the pore. Here we report an experimentally developed hybrid magnetoplasmonic nanopore platform based on bull's-eye geometry that concentrates surface plasmon polaritons into the pore, resulting in significant electric-field enhancement and improved signal readout. The addition of a ferromagnetic layer allows for magnetic tweezing of magneto-plasmonic nanoparticle-tagged molecules, providing active control over their translocation dynamics. Simulations reveal a further boost in enhancement arising from mirror-on-mirror plasmonic coupling between the nanopore and wall-aligned tagged nanoparticles. Together, experimental realization and simulation-guided insights establish a magnetically configurable, plasmonically enhanced nanopore platform that combines signal amplification with controlled translocation for advanced single-molecule sensing and sequencing.

**KEYWORDS:** *Nanopores, plasmonics, single-molecule sequencing, magneto-plasmonics, active control, magnetic tweezing*.

## Introduction

Plasmonic nanopores - sub field of solid-state apertures that incorporate metallic nanostructures—have emerged as a versatile platform for multiplexed biosensing, protein and DNA sequencing, and plasmon-enhanced spectroscopy applications because of their unique ability to combine optical and electrical readouts. [1,2,3,4,5,6,7] Plasmonic nanopores generate strong near-fields, which can function as tweezers capable of trapping nanoparticles or tagged biomolecules near the pore which enable enhanced spectroscopy and sequencing.[8,9,10] Despite significant progress, key challenges persists: excessive fast translocation time of molecules through the nanopore, which limits the interaction time, and the intense local optical fields can cause localized heating and structural instabilities, affecting reproducibility and long-term operation. [11,12,13]

Recent efforts have sought to augment plasmonic systems with active control mechanisms based on DNA based self-assembled techniques, thermal and electrical approaches.[8,9] Magnetism represents another powerful active mechanism for controlling nanoscale motion in plasmonic nanopores. Magneto-plasmonic (MP) nanopores, integrate ferromagnetic (FM) layers and magneto-plasmonic nanoparticles into plasmonic nanopore systems, enable external magnetic-field controlled trapping and active manipulation of tagged molecules.[14,15] This synergy provides a means to regulate the translocation velocity, orientation, and positioning of magnetic or magneto–plasmonic tagged nanoparticles near the pore, while simultaneously exploiting optical resonances for detection, actuation, or local amplification for enhanced spectroscopy.

In this work, we extend these ideas by employing a plasmonic bull's-eye lens—a circular/concentric grating structure that efficiently focuses surface plasmon polaritons into a central nanopore. These bull's-eye structures are renowned for their remarkable extraordinary optical transmission (EOT), which allows light to be confined and concentrated into deeply subwavelength openings.[16-19] In this work, we demonstrated that the combination of high field-focusing capability with magnetic functionality, achieved by incorporating magnetic layer, results in an actively programmable platform for nanoscale confinement, particle manipulation, and improved optical response.

Although photocatalytic metal growth [21] and focused ion beam milling [4,22] have been employed to create plasmonic nanopores with significant electromagnetic confinement at the nanoscale, their magnetic functionality has reamined unexplored. Our group´s previous theoretical word predicted that introducing a thin FM layer either cobalt (Co) or permalloy ($Fe_{80}Ni_{20}$, Py) between thick gold (Au) layers in a nanopore would yield magnetically tunable trapping potential and localized nanocavities with enhanced optical fields.[14]

In this work, we present a hybrid magneto-plasmonic nanopore platform integrating vastly improved optical field enhancement and magnetic trapping functionality. The device employs a bull´s eye gold architecture incorporating an FM layer, which functions as electromagnetic lens that enhances the local field within the nanopore, while providing a magnetic landscape for magnetic particle trapping. Optical and magneto-optical Kerr effect (MOKE) measurements, supported by electromagnetic (Lumerical Finite Differential Time Domain -FDTD) and micromagnetic (Mumax3) simulations, demonstrated enhanced electromagnetic fields in the central pore that surpass those of previously proposed hybrid systems, together with magnetic configurations capable of generating localized magnetic traps within the nanopore.

To evaluate trapping performance, we modelled non-toxic magneto-plasmonic (MP) Janus nanoparticles (100 nm polystyrene nanodomes) with alternating magnetic and plasmonic layers (Fe/Au or Co/Au). These particles offer facile functionalization, colloidal stability, and optical and magnetic anisotropy, enabling magnetophoretic manipulation of multiple particles.[15, 26-30] Their combined magnetic actuation and plasmonic response make them well suited for integration into magnetically configured nanopores for particle manipulation, molecular trapping, and single-particle spectroscopy. For biomolecule-functionalized MP Janus nanoparticles, the designed nanopore system is predicted to exert magnetic trapping forces of ~30–35 nN.

Our initial optical simulations using Lumerical FDTD showed that structures made on pure Au films and Au/Co/Au tri-layers exhibited only minor differences in optical responses, with the addition of the thin FM layer causing a slight decrease in reflection (supplementary Figure S1). Additionally, simulations showed that pure Au bull's-eye structures had a few percent higher electric field enhancement ($E/E_0$) at the nanopore center than that of the Au/Co/Au bull's-eye structure (Figure S2). These results indicate that although the magnetic layer is somewhat detrimental to the system's optical performance, the current nanopore-based design shows an $E/E_0$ of roughly 180, which is about 17 times higher than as described in Maccaferri N et al (2021) work.[14] The improvement in the performance becomes even more striking when a core-shell particle with thicknesses like Maccaferri N et al (2021)[14] is introduced near the nanopore, the $E/E_0$ at gaps of 2 nm and 0.5 nm from the pore wall reach values of around 800-1000 (Figure S3), while the referred paper reports only ~20-30.

## Results and Discussion

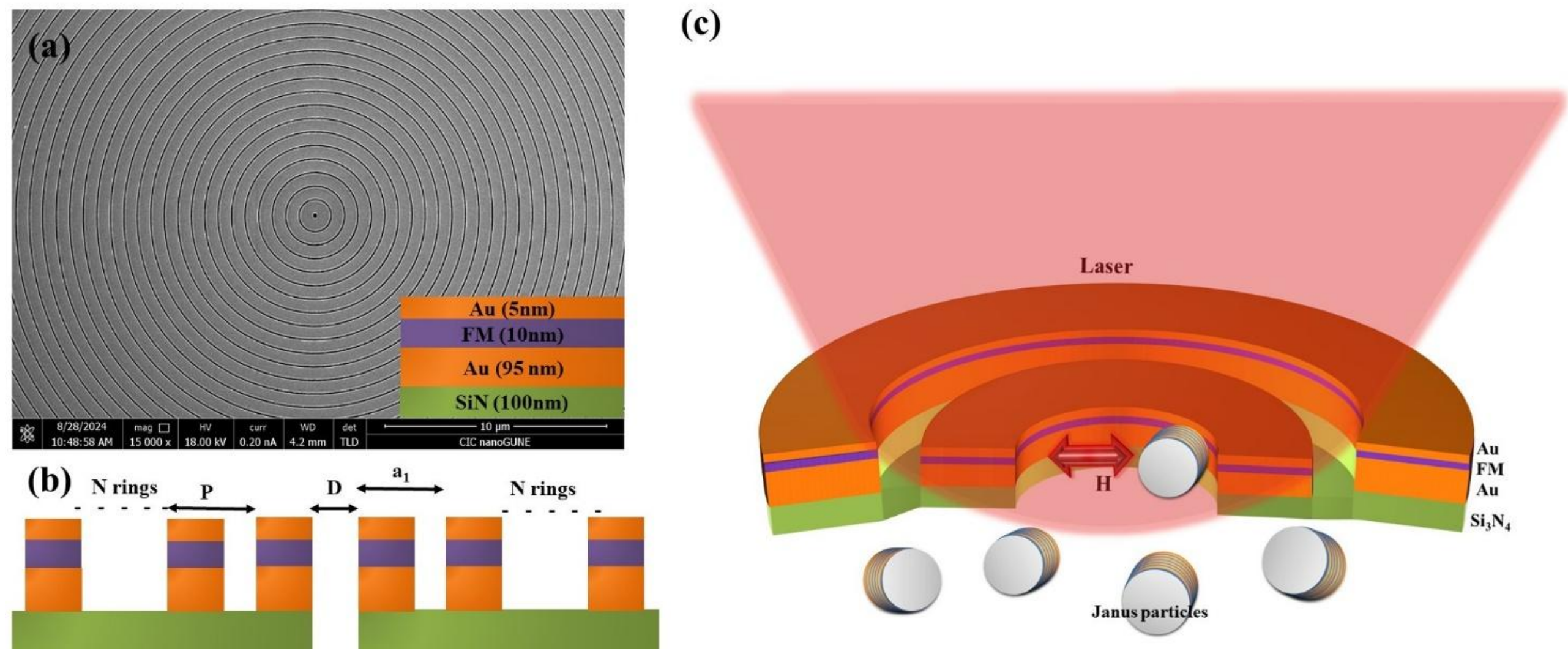

**Figure 1.** SEM image of the concentric ring structure with nanopore at the center with inset showing multilayer magnetoplasmonic thin films (b) cross sectional image of the concentric structure with N rings of varied periodicity (P), Groove width (GW), Groove depth (GD), a- edge of the pore to first groove (c) schematic of the proposed experiment.

Reliable optical and magnetic performances in magnetoplasmonic nanopores require precise control over film thickness and surface roughness. To minimize roughness and optical losses, Au/Co/(Au) tri-layers were deposited using moderate evaporation rates (<1 Å $s^{-1}$). SEM, FIB cross-sectioning, and AFM were used to corroborate the smooth surfaces and clearly defined layer thicknesses of the film (Figure S4 & S5). To reduce substrate effects, a total multilayer thickness of about 110 nm was chosen based on the close agreement between optical responses computed using the T-matrix approach and experimental results (Figure S6). Systems with final thicknesses are shown in inset figure 1(a).

Magneto-plasmonic nanopores were fabricated by focused ion beam (FIB) milling of tri-layer Au/FM/Au films deposited on suspended $Si_3N_4$ membranes. The magnetoplasmonic stack consisted of Au (100 nm)/FM – either Co or Py (5–10 nm)/Au (5 nm), with a 5 nm Ti adhesion layer. Periodic circular gratings (also referred to as vortex rings or bullseye structures) of different geometries with a single nanopore at the center were fabricated by focused ion beam (FIB) milling while varying the beam currents and voltages (SEM image in Figure 1(a)). The groove width (GW) was systematically changed from 50 to 200 nm, while the pore diameter (D = 200 nm), grating period (P = 600 nm), and groove depth (≤110 nm) remained fixed. The distance between the edge of the central pore and the first groove edge (a1) was maintained equal to the pitch (a1 = P). Schematic of the cross-section and the proposed experimental setup are shown in figure 1(b) & 1(c).

Surface plasmon polaritons (SPPs) wavelength supported at a metal–dielectric interface are characterized by the wavevector

$$k_{\text{SPP}} = \frac{2\pi}{\lambda_0}\sqrt{\frac{\varepsilon_m \varepsilon_d}{\varepsilon_m + \varepsilon_d}}$$

where $\varepsilon_m$ and $\varepsilon_d$ are the dielectric functions of the metal and dielectric, respectively, and $\lambda_0$ is the free-space wavelength.

Because the square-root term is greater than unity, the resulting $k_{\text{SPP}}$ is always larger than the in-plane momentum of a free-space photon $k_0 \sin\theta$. This dispersion relation therefore makes it clear that SPPs cannot be excited directly by simply illuminating a flat metal–dielectric interface, as the incident light lacks the required in-plane momentum for phase matching.

To overcome this momentum mismatch, a periodically corrugated metallic surface can supply the additional in-plane momentum required to excite SPPs. In such structures, the grooves act as a diffraction grating that modifies the in-plane momentum of the incident light to

$$k_{\mathrm{SPP}} = k_0 \sin\ \theta + \frac{2\pi l}{p}$$

where $k_0$ is the is the wavevector of the incident light, $p$ is the grating period, $\theta$ is the incidence angle, and $l$ is the diffraction order.

Once excited, the SPPs propagate radially along the circular grooves and interfere constructively at the central aperture, effectively funneling light into a region far smaller than the incident wavelength. The constructive interference that enables this enhancement occurs when the SPPs launched from each groove arrive at the aperture with a phase satisfying the standing-wave condition $\boldsymbol{2k_{\mathrm{SPP}}a_1 = 2\pi m}$, or equivalently $\boldsymbol{a_1 \approx m\lambda_{\mathrm{SPP}}/2}$, where $\boldsymbol{a_1}$ is the distance between the aperture and the first groove, $\boldsymbol{\lambda_{\mathrm{SPP}}}$ is the SPP wavelength and $\boldsymbol{m}$ is an integer representing the number of half SPP wavelengths supported in that cavity region. This plasmon-assisted concentration mechanism can enhance the transmitted power through the aperture by orders of magnitude compared with an isolated hole. Numerous studies have shown that the magnitude of this enhancement is highly sensitive to geometric parameters such as groove depth, width, periodicity, and the groove-to-aperture distance, as these factors govern both the amplitude and phase of the SPPs converging at the center.[16-19]

The plasmonic response of the fabricated bull's-eye structures were analyzed using optical reflection spectra (Figure 2(a)). For first order coupling $(\boldsymbol{l = 1})$ and grating period $\boldsymbol{p = 600\ \mathrm{nm}}$, the surface plasmon polariton (SPP) wavelength predicted by the momentum-matching relation is ~670 nm whereas experimentally this SPP-driven resonance appears red-shifted wavelength to ~740 nm. Additional features can be seen at 630 nm and 850 nm. FDTD simulations were performed to understand the behavior. The influence of groove depth (with groove width fixed at 100 nm and period fixed at 600 nm) is shown in figure 2(b). As shown in Figure 2(c), SPP excitation is present even for shallow grooves, but the coupling is very weak at low etching depths. Increasing the groove depth strengthens the interaction between the incident field and the grating, resulting in more efficient SPP generation and a red shift in the resonance wavelength due to longer effective optical path; groove width causes only a minor red shift (Figure S7). Simulations reveal an additional feature near 750 nm as seen in figure 2(b), which arises from the excitation of a collective groove resonance that can eventually hybridize with the groove-mediated SPP mode.[18] while experimentally, this collective resonance appears further red-shifted, around 850 nm as seen in figure 2(a), and dominating the spectrum likely due to uncontrolled variations in the actual groove depth during fabrication. In one of our fabricated structures, we also observed (Figure S8), the hybridization between these two plasmonic responses. Up to this point, our observations were consistent with trends reported in bull's-eye literature. However, the additional feature around 630 nm in our measurements was unexpected—no previous studies report such feature. A key difference is that most reported devices employ Au films thicker than 250 nm, with groove depths far from the metal–substrate interface. In contrast, our structure employs a 110 nm Au film and the grooves depths of ~ 90-110 nm, bringing grooves closer to the substrate which leads to evanescent coupling of Au/Air SPP and Au/SiN interface. A detailed discussion of the SPP penetration depth and the emergence of the hybridized mode as a function of groove depth is provided in the Supplementary Information (Figure. S9). [31]

Although experimental observations follow the trends predicted by simulations, the different spectrum measured for GW = 100 ± 25 nm - characterized by a broad, red-shifted feature, can likely be attributed to a thinner Au layer at the bottom of the groove (i.e., a slightly larger GD exceeding 50 nm) produced during FIB milling of the wider groove. This is consistent with the high sensitivity of the spectral response to the Au thickness at depths beyond ~60 nm. Additional discrepancies can reasonably be attributed to fabrication-related variations such as uncontrolled groove-width fluctuations, spacing deviations, and surface roughness and limited groove depth control during FIB milling as well as, local gallium implantation, redeposition, and tapering of the groove sidewalls slightly modify the effective dielectric environment and the plasmonic response.

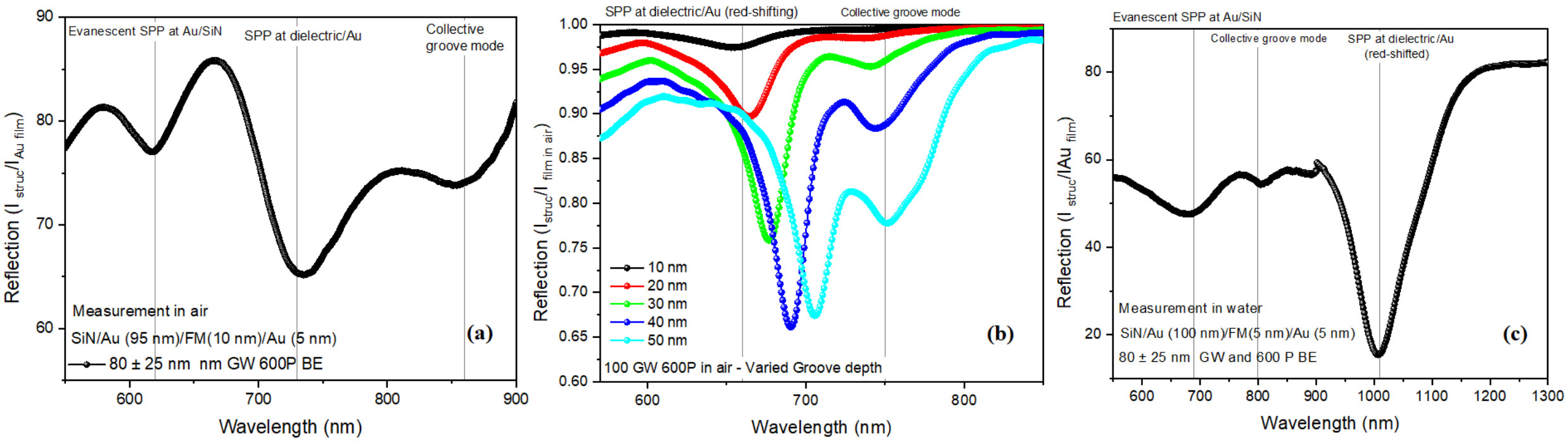

**Figure 2.** Optical response of the concentric ring structure corresponding to the 600P and 80± 25 nm groove depth measured in (a) air and (c) in water. Lumerical FDTD simulations in air, showing red shift upon varying groove depth at fixed groove width (b)

We further examined the influence of superstrate on the optical response, since the final experiments are performed in water with nanoparticles (Figure 2(c)). Replacing air with water, resonances exhibited red shifts, but with significantly different magnitudes due to their different modal origins. The feature near ~630 nm corresponds to evanescently coupled mode associated with the Au/SiN interface. This mode resides predominantly on the substrate side of the metal, and its field distribution is governed primarily by the Au–SiN interface, which remains largely unchanged when switched from air to water. Consequently, this resonance shifts moderately from ~630 nm to ~700 nm. This behavior is consistent with the fact that the metal-side penetration depth $\delta_m$ (~25–30 nm) (see Supplementary Information for the expression used to estimate $\boldsymbol{\delta_m}$) depends primarily on the intrinsic optical properties of gold and is only weakly influenced by the refractive index of the superstrate. Strong evanescent coupling is expected when the metal film is sufficiently thin, typically ~50 nm[31] and can further be strengthened by matching the refractive index of superstrate and substrate, which increases the optical symmetry and facilitates more efficient mode hybridization.[32]

In contrast, the resonance near ~750 nm arises from the grating-coupled SPP at the dielectric/Au interface on the top surface. Replacing air (ε≈1) with water (ε≈1.77) strongly modifies the SPP wavevector/confinement by reducing the dielectric-side penetration depth, which scales approximately as [31]

$$\boldsymbol{\delta_d \propto \sqrt{\frac{\varepsilon'_m + \varepsilon_d}{\varepsilon_d^2}}}$$

Where $\boldsymbol{\varepsilon'_m}$, is the real part of the gold permittivity and $\boldsymbol{\varepsilon_d}$ is the dielectric permittivity. Increasing the superstrate permittivity therefore produces a more tightly bound and more confined surface mode at the water/Au interface. Because this SPP resides primarily on the top dielectric side, it is highly sensitive to changes in the surrounding refractive index. As a result, this resonance undergoes a substantial red shift (from ~750 nm to ~1000 nm) and becomes more pronounced when the device is immersed in water, consistent with stronger field localization at the nanopore center. The exact magnitude of the shift is determined by the degree of wetting prior to optical measurements.

Since the goal of this work is to integrate magnetic functionality into the plasmonic nanopore lens and towards this as mentioned, a thin FM layer either Co or Py of 5 - 10 nm thick was inserted between the gold films during the fabrication. It is important to note that this FM layer is located near the top of the film, meaning that the system consists of ferromagnetic rings. Although Co provides strong a high saturation magnetization ($M_s$) and thus strong stray fields, it is susceptible to oxidation when exposed to aqueous or oxygen-rich environments, which can compromise stability. In the nanopore architecture, the Co layer, despite being embedded between Au films but is still partially exposed along the pore sidewalls, where oxidation may occur. To avoid this potential degradation, Py was selected as a more chemically stable alternative for the nanopore design.

Lumerical FDTD showed that Bull's eye structure made of Au films and Au/FM/Au trilayers exhibited only minor differences in their optical response, with the addition of FM layer causing a slight decrease in reflection (Figure S1). The electric field enhancement at the nanopore center is marginally higher for pure Au than for Au/Co/Au and substituting Co with Py gives comparable results (Figures S2, S10), indicating minimal impact of the magnetic layer on optical performance.

After assessing the nanopore's optical response, we now focus on magnetic tweezers by adjusting the magnetic configuration of the ferromagnetic layers within the plasmonic nanopore lens. By embedding a thin ferromagnetic interlayer within a concentric plasmonic nanopore lens, we create nanoscale magnetic tweezers that trap and precisely position magnetic nanoparticles near the pore wall.

The magnetic properties of the concentric ring structures were characterized using longitudinal magneto-optical Kerr effect (MOKE) microscopy. Representative hysteresis loops for Py rings are shown in Figure 3a, measured with an in-plane applied magnetic field. The patterned concentric rings have larger coercive field (i.e., switching field) loops than continuous films, as expected because of spatial confinement. Similar behavior can be seen in cobalt samples (Figure S11). Circular ferromagnetic rings support two stable magnetization states: an onion state with opposing domain walls and a flux-closure vortex state. While ideal rings switch directly between reversed onion states, asymmetries can induce a vortex-like intermediate at low fields. Although not explored here, such vortex states could enable controlled particle de-trapping by eliminating stray fields in the pore, offering potential functionality for future magneto-optical systems.[33-38]

Micromagnetic simulations were performed for cobalt rings using MuMax3 to visualize the magnetization and stray-field distributions. The computational grid consisted of 400 × 400 × 50 cells with a 4 × 4 × 2 $nm^3$ cell size, corresponding to a 1.6 μm × 1.6 μm × 100 nm simulation box. The magnetic region was modeled as a 10 nm film (2x5 nm layers) patterned into two concentric rings surrounding a 200 nm central nanopore, with inner ring ID/OD = 200/700 nm and outer ring ID/OD = 800/1300 nm. The Py layer parameters were saturation magnetization $\boldsymbol{M_s = 8.6 \times 10^5}$ $A \cdot m^{-1}$, exchange stiffness constant $\boldsymbol{A_{ex} = 1.3 \times 10^{-11}}$ $J \cdot m^{-1}$, and magneto-crystalline anisotropy $\boldsymbol{K_c = 0}$. The magnetization was initialized along both the x- and y-axes, saturated under an external field of 30 mT, and then relaxed to 0 mT field. The simulated in-plane magnetization maps (Figures. 3(b–c), left) and corresponding demagnetizing energy density maps (right) reveal localized high-energy regions near the nanopore edge. These regions correspond to concentrations of surface and volume magnetic charges $\mathbf{M} \cdot \mathbf{n}$ and $\mathbf{\nabla} \cdot \mathbf{M}$, respectively. In particular, surface magnetic charge $\mathbf{M} \cdot \mathbf{n}$ at the pore edge generates intense, spatially confined stray fields acting as magnetic tweezers.

These fields define the active trapping zone, enabling nanoparticles to experience significant magnetic forces and achieve stable capture near the pore wall. Additional surface charges are generated in the narrow groove regions (≈80 ± 25 nm wide), whose width is much lesser than particle size considered in this work. As a result, the magnetic tweezing effect within the groove is negligible. In the pore region, however, the pole separation (≈200 nm pore diameter) is large enough for the stray fields to create a spatially accessible trapping landscape capable of capturing nanoparticles.

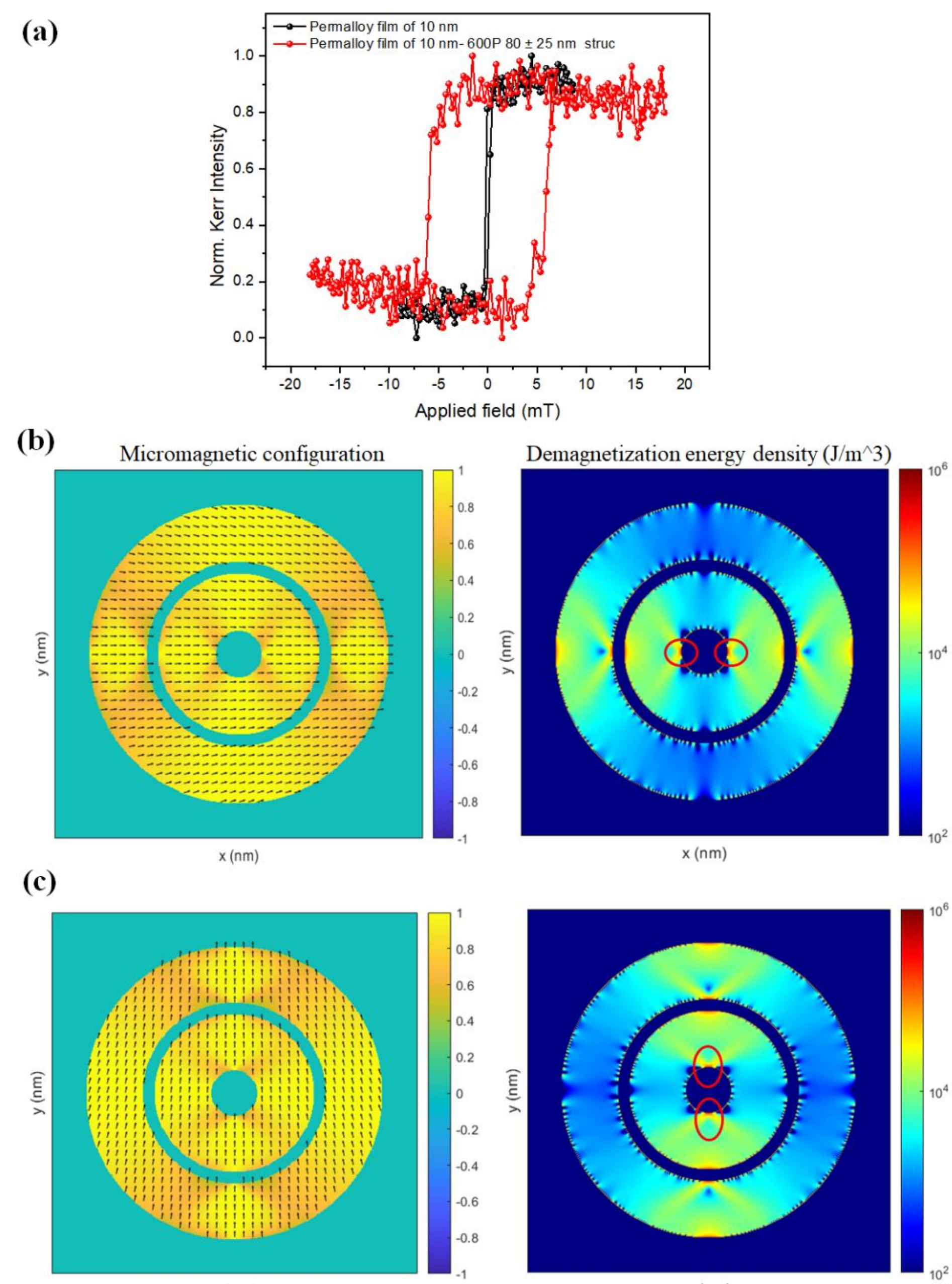

**Figure 3.** Magnetic response measured by magnetooptical Kerr effect microscopy (MOKE) in (a) permalloy rings of 600P and 80 ± 25 nm GW with 200 nm central pore. Micromagnetic simulations correspond to permalloy after saturating and relaxing to remanence states in x and y directions (b) and (c) (with magnetic configurations (left) and demagnetization energy density(right) with tweezers location shown in red circles). The MOKE loops represent the magnetic response of the real samples, which contain many rings, whereas the simulations correspond only to the central portion of the structure.

The simulations also provide the magnetic stray field $\mathbf{H}$ generated by the induced magnetic tweezers, from which the magnetic field gradient $\nabla |\mathbf{H}|$ was evaluated directly from the MuMax3 demagnetizing field. The calculated 2D maps of $|\nabla\mathbf{H}|$ (Figure 4a) reveal intense gradients localized at the nanopore edges, reaching magnitudes on the order of 10^8 T/m.

Magneto-plasmonic nanopore membranes are designed to work in tandem with magneto-plasmonic nanoparticle (NP), allowing simultaneous optical and magnetic control of nanoscale objects. In this study, we investigated magneto-plasmonic Janus nanoparticles, following the design of Li et al (2018).[26,27] These nanodomes are a new type of multifunctional hybrid nanostructure designed for optical, magnetic, and biomedical applications. The particles comprise of 100 nm fluorescent polystyrene (PS) core, half-coated with a multilayer stack of (Co (1 nm)/Au (6 nm)) ×5, and capped by an additional 6 nm Au layer, forming a hemispherical metallic shell shown in Figure 4(b). This asymmetric configuration combines ferromagnetic activity and plasmonic resonance, The Co layers provide field-responsive magnetic mobility and, owing to their reduced thickness results in perpendicular magnetic anisotropy (PMA),[39] whereas the Au layers maintain strong plasmonic resonances.[26] Unlike in Au/Co/Au nanopore where pore walls are partially exposed to external environment, Janus particles exhibit high chemical stability in aqueous environments because their magnetic Co layers are fully encapsulated by a continuous external Au shell, preventing direct exposure to the medium. Together these properties yield non-toxic, colloidally stable nanoparticles that show strong magneto-chromic modulation, with optical spectra influenced by both magnetic field orientation and light polarization, making them ideal functional labels for the magneto-plasmonic nanopore platform.

Further, magnetic force acting on a nanoparticle computed (assuming the particles are saturated and aligned parallel to the magnetic stray field) as follows

$$\mathbf{F} = \boldsymbol{\mu_0}(\widehat{\mathbf{m}} \cdot \boldsymbol{\nabla})\mathbf{H},$$

where $\widehat{\mathbf{m}} = \boldsymbol{M}(\boldsymbol{H})\, \boldsymbol{V}_{\mathbf{Co}}\, \hat{\mathbf{h}}$, and $\hat{\mathbf{h}}$ is the unit vector parallel to the local stray field $\mathbf{H}$. Unlike the superparamagnetic magnetite nanoparticle considered in the previous work, we investigated an anisotropic Janus particle using the experimentally measured $M(H)/M_s$ taken from Li et al 2018[26] (supplementary Fig. S12). The magnetization response indicates that the magnetization of the Janus particles will be saturated along H near the pore wall, so that the trapping force can be calculated as $F=\mu_0 m_s \nabla H$, with $m_s = M_s{*}V_{Co}$. For the force calculations, saturation magnetization was taken as $\mathbf{M}_{\boldsymbol{s}} = \mathbf{1.4}\boldsymbol{X}\mathbf{10^6}\ \mathbf{A} \cdot \mathbf{m^{-1}}$, corresponding to the standard value of Co.

The simulated 2D magnetic force maps (Figure. 4(a)) also show that the Janus particle experiences a maximum trapping force of ≈ 30 nN at the Au/Py/Au nanopore edge, where the magnetic field gradient reaches ~$\mathbf{10^8}\ \mathbf{T} \cdot \mathbf{m^{-1}}$ (Figure 4(a)). Simulations of Au/Co/Au nanopore architecture produced a substantially higher maximum force of ~32 nN (Figure S13).

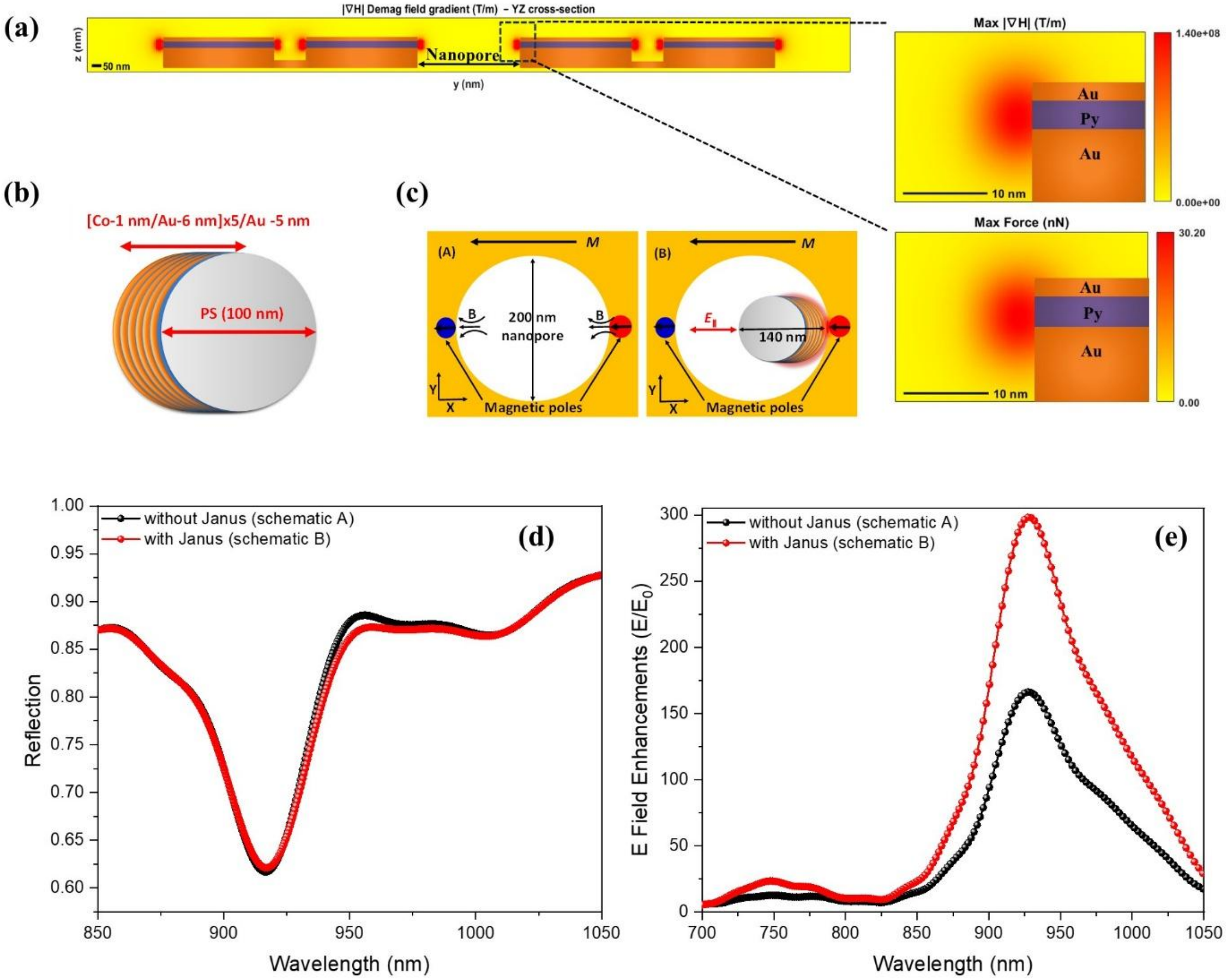

**Figure.4.** (a) Mumax3 simulations of the gradient field in a Au/Py/Au nanopore and its corresponding calculated force acting on a janus particle. Schematic of Janus particles with composition (b), (c) Magnetic tweezer system illustrating without a particle (A) and with resulting trapping of Janus particles (B) in a nanopore aligned parallel to the x-polarized source (x-y plane of the film). (d) Lumerical FDTD – simulated reflection spectra of the plasmonic nanopore and its corresponding (e) near electric field enhancement for particles positioned at 2 nm from the pore wall

Upon magnetic trapping, the particle is expected to orient with its metallic half-shell facing the pore wall. To achieve strong field enhancement exploiting the plasmonic nanogap, E field must be aligned along the Janus particle axis, defined as the axis normal to the interface between the two hemispheres. Simulated reflection spectra (Fig. 4(d)) show only tiny variations, indicating that the presence of the particle does not significantly modify the far-field response (reflection) of the lens. In contrast, significant enhancement effects $E/E_0$ are observed in the near field. When the Janus axis is aligned parallel to the polarization, the $E/E_0$ reaches ~300—approximately twice that of the nanopore alone (Fig. 4(e)). This behavior is consistent with the trapping geometry expected under perpendicular magnetic anisotropy (PMA) in the Co/Au Janus particles, where the gold-coated hemisphere preferentially aligns toward the ferromagnetic nanopore wall. These confinement values remain effectively unchanged when the cobalt in the nanopore system is replaced with Permalloy (supplementary Fig. S14).

To benchmark our findings, we also simulated a core–shell nanoparticle with dimensions comparable to those studied by Maccaferri N et al (2021).[14] In our magneto-plasmonic nanopore architecture, $E/E_0$ (Figure S3) at 2 nm and 0.5 nm particle–wall gaps reaches ~800–1000, whereas the referenced work reports only ~20–30. This highlights the substantially stronger optical enhancement achieved by the present design.

In conclusion, we have developed a hybrid magneto-plasmonic nanopore technology that blends a Bull's-eye lensing design with externally configurable, magnetic-field-controlled trapping. The concentric groove design efficiently concentrates light into the nanopore, generating strong near-field enhancement that supports sensitive, nanoscale optical investigation. By embedding a thin ferromagnetic layer within the Au rings, the system maintains high optical confinement while introducing substantial stray-field gradients ($>10^8$ $T \cdot m^{-1}$), enabling nanoscale magnetic tweezers capable of exerting trapping forces up to ~30 nN on magneto-plasmonic Janus particles.

Electromagnetic simulations further reveal large, orientation-dependent enhancements: $E/E_0 \approx 300$ (at a gap of 2 nm from pore wall) when the Janus axis aligns with the incident polarization, and up to ~800–1000 when the core-shell particle is positioned 0.5–2 nm from the pore wall—far exceeding previously reported values ~20–30. These enhanced fields, combined with active magnetic configuration, enable precise manipulation of particle position and orientation, providing a route to improved optical sensitivity during capture, confinement, and translocation.

Together, these results demonstrate the first magnetically reconfigurable plasmonic lens nanopore capable of simultaneous optical focusing and magnetic tweezing. By merging strong near-field enhancement with external magnetic control, this architecture enables active, real-time manipulation and investigation of nanoscale particles, establishing a versatile foundation for magnetically assisted plasmonic tweezers, nanoscale actuation, and high-sensitivity, multimodal lab-on-a-chip systems.

ASSOCIATED CONTENT

Supporting Information.

(1-2). Simulated Optical response (reflection) and Electric field enhancement showing comparison of a Au -only device with that of an identical Au/Co/Au trilayer respectively; (3) Electric field enhancement with core shell particle in a bull's eye nanopore made on Au/Co/Au tri-layer. (4-5) Scanning electron microscopy images (SEM) and atomic force microscopy images (AFM) of Au/Co/Au trilayer films respectively; (6) Theoretical (T-matrix) optical response (reflection) of as grown multilayer films on Silicon substrate (7) Lumerical FDTD simulations in air, showing red shift upon varying groove width (GW) at fixed groove depth (GD) (8) Experimental optical response (measured in air) of the concentric ring structure corresponding to the 600P and 100 (+/- 25 nm) groove depth made on Au/Co/Au measured in air (9) Lumerical FDTD simulations showing Evanescent SPP (of Air/Au) – interference with $Si_3N_4$ at deeper etchs ; Discussion S9 (10) Comparison of the simulated electric-field enhancement for Au/Co/Au and Au/Py/Au nanopores with a trapped Janus particle under x-polarized excitation at a 2 nm nanogap from pore wall. (11) Magnetic response measured by magnetooptical Kerr effect microscopy (MOKE) of concentric ring structure (600 P and 80 (+/-25) nm GW structure) made on cobalt of Au (95 nm) /Co (10 nm)/Au (5 nm) film (12) Room temperature out-of-plane hysteresis loop for the Co/Au Janus particles measured by vibrating sample magnetometer (VSM) taken from Li et al 2018 (13) Mumax3 simulation of the gradient field in a Au/Co/Au nanopore and its corresponding calculated force acting on a janus particle

AUTHOR INFORMATION

Corresponding Authors

nr.sanamreddy@nanogune.eu

p.vavassori@nanogune.eu

Authors Contributions

NS conceived and developed the fabrication method, performed the numerical simulations, and wrote the manuscript. PV provided supervision and oversight throughout the work

ACKNOWLEDGMENT

The authors thank the European Union under the Horizon 2020 Program, HORIZON-MSCA-DN-2022: DYNAMO, grant Agreement 101072818.